\documentclass[prd,twocolumn,floatfix,amsmath,nofootinbib,amssymb,floatfix]{revtex4}
\usepackage{graphicx,color,dcolumn,booktabs,bm,multirow}
\usepackage{longtable,lscape}
\usepackage{txfonts}
\usepackage{overpic}
\usepackage{amssymb}
\usepackage{array}
\usepackage{indentfirst}
\usepackage{feynmf}   
\usepackage{bbding}
\usepackage{slashed}  
\usepackage{cases}
\usepackage{color}
\usepackage{multirow}
\usepackage{epstopdf}
\usepackage{tabularx}
\usepackage{graphicx,color,dcolumn,booktabs,bm}
\usepackage[compat=1.1.0]{tikz-feynman}
\usepackage[colorlinks,
citecolor=blue,
anchorcolor=red,
menucolor=red,
linkcolor=red,
filecolor=red,
runcolor=red,
urlcolor=blue,
frenchlinks=red]{hyperref}

\def\X0{X_0(2900)}

\begin{document}

\title{$X_0(2900)$ and its spin partners productions in the $B^+$ decay process }
\author{Zhuo Yu$^{1}$}
\author{Qi Wu$^{2}$}
\author{Dian-Yong Chen$^{1,3}$\footnote{Corresponding author}}\email{chendy@seu.edu.cn}
\affiliation{$^1$ School of Physics, Southeast University, Nanjing 210094, China}
\affiliation{$^2$Institute of Particle and Nuclear Physics, Henan Normal University, Xinxiang 453007, China}
\affiliation{$^3$ Lanzhou Center for Theoretical Physics, Lanzhou University, Lanzhou 730000, China}
\date{\today}

\begin{abstract}
In the present work, we investigate $X_0(2900)$ and its spin partners  productions in the $B^+$ decay processes, where $X_0(2900)$ is regarded as a $\bar{D}^\ast K^\ast$ molecular state. We infer that the initial $B^+$ meson and the final $D^+ X_0(2900)$  are connected through $D_{s1}^+(2460) \bar{D}^{\ast 0} $ mesons by exchanging a $K^{\ast 0}$ meson. The contributions from such a meson loop to the process $B^+ \to D^+ X_0(2900)$ are evaluated and our estimations indicate that the measured branching fraction of $B^+\to D^+ X_0(2900)$ could be well reproduced in a reasonable model parameter range, which indicates that $X_0(2900)$ could be interpreted as a $\bar{D}^\ast K^\ast $ molecular state. In addition, the production processes $B^+ \to D^+ \tilde{X}_{1,2}$, $B^+ \to D^{\ast +} X_0(2900)$, and $B^+ \to D^{\ast +}\tilde{X}_{1,2}$ are investigated, and we find that the fit fraction of $\tilde{X}_1$ in the process $B^+ \to D^{\ast +} D^{\ast -} K^+$ is aobut $6\%$, which should be accessible experimentally.
\end{abstract}
\maketitle

\section{introduction}
\label{sec:introduction}

Over the past two decades, there are a growing number of new  hadron states  reported by Belle, BESIII, and LHCb collaborations (for further details, refer to the reviews~\cite{Guo:2017jvc,Olsen:2017bmm, Brambilla:2019esw,Ali:2017jda,Guo:2019twa,Liu:2013waa,Dong:2017gaw,Liu:2019zoy}). Among these exotic candidates, the majority have been observed in hidden charm channels, exemplified by $X(3872)$~\cite{Belle:2003nnu} and $Z_c(3900)/Z_c(4020)$\cite{BESIII:2013ouc,BESIII:2013mhi}, which are referred to as charmonium-like state. 

In addition to the charmonium-like states, another intriguing category of new hadron states is the fully open flavor states. As typical examples of this kind of new hadron states, $X_{0,1}(2900)$ were observed by the LHCb Collaboration in the $D^-K^+$ invariant mass distributions of the decay process $B^+ \to D^+D^-K^+$ in 2020 \cite{LHCb:2020pxc,LHCb:2020bls}. The helicity amplitudes analysis indicated that the spins of these two structures were 0 and 1, respectively. Consequently, the $J^P$ quantum numbers of $X_0(2900)$ and $X_1(2900)$ are $0^+$ and $1^-$, respectively. The measured mass and width of $X_0(2900)$ were
	\begin{equation}
		\begin{split}
		m_0 &= (2866\pm7\pm2) \, {\rm MeV}, \\
		\Gamma_0 &=  (57\pm12\pm4) \, {\rm MeV},
		\end{split}
	\end{equation}
while the ones of $X_1(2900)$ were reported to be, 
	\begin{equation}
	\begin{split}
		m_1 &= (2904\pm5\pm1) \, {\rm MeV}, \\
		\Gamma_1 &=(110\pm11\pm4) \, {\rm MeV}.
	\end{split}
\end{equation}

From the observed channel of $X_{0,1}(2900)$, it is apparent that the quark compositions of $X_{0,1}(2900)$ are $ud\bar{s}\bar{c}$, indicating that $X_{0,1}(2900)$ are fully open flavor exotic states. Consequently, investigations into their properties hold significant potential for enhancing our understanding on the non-perturbative behaviors of the strong interaction in QCD. Inspired by the observations of $X_{0,1}(2900)$, various tetraquark interpretations have been proposed in different frameworks. The estimations in the framework of the improved chromo-magnetic interaction model indicated that $X_0(2900)$ could be a compact tetraquark state composed of $ud\bar{s}\bar{c}$~\cite{Guo:2021mja}. The mass of the diquark-antidiquark states were estimated by using the QCD sum rule and the authors in Refs.~\cite{Zhang:2020oze, Wang:2020xyc} found that $X_0(2900)$ could be assigned as an axialvector-diquark-axialvector-antidiquark type tetraquark states. By considering the SU(3) flavor symmetry, the authors in Ref.~\cite{He:2020jna} interpreted the $X_0(2900)$ as the radial excited tetraquark state with $J^P=0^+$. In Ref.~\cite{Wang:2020prk}, the compact $\bar{c}\bar{s}qq$ tetraquark states have been systematically estimated in the quark model with the Coulomb, the linear confinement, and the hyperfine interaction, and the estimated mass and width of the ground state with $I(J^P)=1(0^+)$ are consistent with that of $X_0(2900)$. some estimations challenge the tetraquark interpretations of $X_0(2900)$. For instance, the estimations in the extended relativized quark model indicate that the mass of ground state of compact $\bar{c}\bar{s}qq$ tetraquark states with $J^P=0^+$ is approximately 100 MeV below the one of $X_0(2900)$~\cite{Lu:2020qmp}. Additionally, the spectroscopic parameters of either axial-vector and scalar diquark-antidiquark derived by employing the QCD two-point sum rule method are inconsistent with those of $X_0(2900)$ either~\cite{Agaev:2022eeh}.

Furthermore, the measured mass of $X_{0,1}(2900)$ are in close to the $D^\ast K^{\ast}$ threshold, suggesting that these two states could be good candidates of the molecular states composed of $\bar{D}^\ast {K}^\ast$. Considering the $J^P$ quantum numbers of $X_{0,1}(2900)$, one can find that $X_0(2900)$ and $X_1(2900)$ could be a $S-$wave and $P-$wave $\bar{D}^\ast K^\ast$ molecular state, respectively. Before the experimental observation of the $X_{0,1}(2900)$, the authors in Ref.~\cite{Molina:2010tx} predicted the existence of $D^\ast \bar{K}^\ast$ molecule with $J^P=0^+$ within the hidden-gauge formalism. Subsequent to the  observation, the $D^\ast \bar{K}^\ast$ molecular interpretations have been further discussed by various groups. For instance, the estimations within the Bethe-Salpeter framework suggest that $X_0(2900)$ could be identified as a $\bar{D}^\ast K^\ast $ molecular state with $I(J^P)=0(0^+)$~\cite{Ke:2022ocs, Kong:2021ohg}. Similarly,  investigations within the chiral effective field theory also supported the $\bar{D}^\ast K^\ast$ molecular interpretations of $X_0(2900)$, while $X_1(2900)$ was considered as the $P$-wave excitation of the ground-state $X_0(2900)$~\cite{Wang:2021lwy}. By using the QCD sum rule methods, the mass spectroscopy, decay constants of the $\bar{D}^\ast K^\ast$ molecular state had been investigated~\cite{Chen:2021erj, Agaev:2020nrc, Chen:2020aos, Mutuk:2020igv}, which indicate that $X_0(2900)$ could be a $\bar{D}^\ast K^\ast$ molecule with $J^P=0^+$, while the $X_1(2900)$ could be considered as the $P$ wave $\bar{c}\bar{s}qq$ tetraquark with $J^P=1^-$~\cite{Chen:2020aos}. In Ref.~\cite{Liu:2020nil}, the $D^\ast$ and $\bar{K}^\ast$ interaction was explored in the one-boson-exchange model, the mass of $X_0(2900)$ could be reproduced, however, $X_1(2900)$ could not be considered as a $\bar{D}^\ast K^\ast$ molecular state with $J^P=1^-$ since $\bar{D}^\ast K^\ast$ potential is repulsive in the case of $P$-wave. In addition, by using the effective Lagrangian approach, the authors in Refs. ~\cite{Xiao:2020ltm, Huang:2020ptc} estimated the decay width of $X_0(2900)$, which is in agreement with the experimental measurement within the uncertainty of the model. Besides the tetraquark and molecular interpretations, the kinematical cusp effect ~\cite{Swanson:2021fqw,Burns:2020xne}, could also provide insights into understanding these states.

Besides the resonance parameters and decay properties, the production processes are also crucial in decoding the nature of $X_0(2900)$, which sparked numerous theoretical predictions using various approaches. For example, the cross sections for $K^+p \to \Sigma_c^{++}X_0(2900)$~\cite{Lin:2022eau} and decay widths of $\Lambda_b \to \Sigma_c^{0,(++)}X_{0,1}^{'0,(--)}$~\cite{Hsiao:2021tyq} have been investigated. On the experimental side, the fit fraction of $X_0(2900)$ in the process $B^+ \to D^+ D^- K^+$ is measured to be $(5.6\pm 1.4 \pm 0.5)\%$. The branching faction of $B^+ \to D^+ D^- K^+$ is $(2.2 \pm 0.7)\times 10^{-4}$, thus the branching fraction of the two body cascade decay process $B^+ \to D^+ X_0(2900) \to D^+ (D^- K^+)$ is measured to be $(1.2 \pm 0.5) \times 10^{-5}$. The estimations in Refs.~\cite{Xiao:2020ltm,Huang:2020ptc} indicated that the $X_0(2900)$ is dominantly decay into $\bar{D}K$ channel, where $X_0(2900)$ are considered as a $\bar{D}^\ast K^\ast $ molecular state. Considering the isospin symmetry, one can find the branching fraction of $X_0(2900)\to D^- K^+$ is about $50\%$, thus, one can approximately conclude,
\begin{eqnarray}
	\mathcal{B}[B^+ \to D^+ X_0(2900)] \approx (2.4\pm 1.0) \times 10^{-5}.\label{Eq:BRexp}
\end{eqnarray} 
Similar to the process $B^+ \to D^+ D^- K^+$, one can also try to search $X_0(2900)$ in the the process $B^+ \to D^{\ast +} D^- K^+$. The investigations of the production mechanism of $X_0 (2900)$ in the $B^+$ decay would go a long way towards revealing its nature.

\begin{table}
\centering
\caption{The branching ratios of the relevant processes, where $X_0(2900)$, $\tilde{X}_{1,2}$ may be observed. \label{Tab:BR}}
\renewcommand\arraystretch{1.5}
\begin{tabular}{p{2cm}<\centering p{3.0cm}<\centering p{3cm}<\centering}
\toprule[1pt]
State & Process & Branching Ratio \cite{ParticleDataGroup:2020ssz} \\
\midrule[1pt]
\multirow{2}{*}{$X_0(2900)$} & $B^+ \to D^+ D^- K^+$ & $(2.2\pm 0.7)\times 10^{-4}$\\
    &$B^+ \to D^{\ast+} D^- K^+$ & $(6.3 \pm 1.1)\times 10^{-3}$\\
\multirow{2}{*}{$\tilde{X}_{1,2}$} & $B^+ \to D^+ D^{\ast-} K^+$ & $(6.0\pm 1.3)\times 10^{-4}$\\
    &$B^+ \to D^{\ast+} D^{\ast-} K^+$ & $(1.32 \pm 0.18)\times 10^{-3}$\\    
\bottomrule[1pt]	
\end{tabular}
\end{table}

Moreover, in the molecular frame, $X_0(2900)$ is regarded as an $S$-wave $\bar{D}^\ast K^\ast $ molecular state. In addition to  the $0^+$ state, corresponding to $X_0(2900)$, the $J^P$ quantum numbers of the $S-$wave $\bar{D}^\ast K^\ast $ system could also be $1^+$ and $2^+$. In Ref.~\cite{Hu:2020mxp}, with the assumption that $X_0(2900)$ is a $I(J^P)=0(0^+)$ $\bar{D}^\ast K^\ast$ hadronic molecule, the potential of $\bar{D}^\ast K^\ast$ interactions could be determined, and $\bar{D}^\ast K^\ast$ molecular states with $I(J^P)=0(1^+)$ and $I(J^P)=0(2^+)$ were predicted. While the estimations in the one-boson-exchange model respecting heavy quark spin symmetry indicated that  $X_0(2900)$ could be a $\bar{D}^\ast K^\ast$ hadronic molecule with $I(J^P)=0(0^+)$, and an additional bound state with $I(J^P)=0(1^+)$ was predicted, while the one with $I(J^P)=0(2^+)$ might be unbounded~\cite{Liu:2020nil}. Similar conclusion had been drawn in Ref.~\cite{He:2020btl} but the one with $I(J^P)=0(2^+)$ could be a virtual state. By fine tuning the model parameters to reproduce exactly the mass and width of the $X_0(2900)$ in the local hidden gauge formalism, two more states with $J^P=1^+$ and $2^+$ were reported in Ref.~\cite{Molina:2020hde}, and these two additional states have been proposed to be searched in the processes $\bar{B}^0 \to K^{\ast 0} D^{\ast +} K^-$~\cite{Dai:2022htx} and $B^+ \to D^+ D^- K^+$~\cite{Bayar:2022wbx}. From the above literatures, one can conclude that the spin partner of $X_0(2900)$ should exist if one considering $X_0(2900)$ as a $\bar{D}^\ast K^\ast$ hadronic molecule. Thus, searching for its spin partners should also shield light on molecular nature of $X_0(2900)$. Hereafter, we use $\tilde{X}_1$ and $\tilde{X}_2$ refer to the spin partner of $X_0(2900)$ with $J=1$ and $J=2$, respectively. As indicated in Ref.~\cite{Xiao:2020ltm}, $\tilde{X}_1$ and $\tilde{X}_2$ should dominantly decay into $D^\ast K$, consequently, one can expect searching $\tilde{X}_1$ and $\tilde{X}_2$ in the processes $B^+ \to D^{(\ast) +} D^{\ast -} K^+$. In Table~\ref{Tab:BR}, we collected the branching ratios of the relevant processes, where $X_0(2900)$, $\tilde{X}_{1,2}$ may be observed.

This work is organized as follows. After introduction,  we present our analysis of the production mechanisms working in the processes $B^+ \to D^{(\ast) +} X_0(2900)$, and $B^+ \to D^{(\ast) +} \tilde{X}_{J},\ (J=1,2)$, the branching fractions of these processes are estimated in Section \ref{Sec:Method}. The numerical results and related discussions are presented in Section~\ref{Sec:Num}, and the last section is devoted to a short summary.

\section{$X_0(2900)$ and its spin partners Productions in the $B$ decay process}
\label{Sec:Method}

\begin{figure}[t]
	\centering
	\includegraphics[scale=0.8]{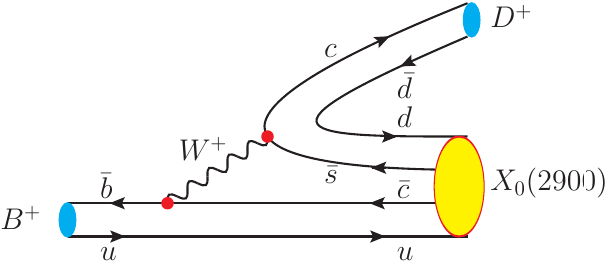}
    \caption{(Color online.) Diagram contributing to $B^+ \to D^+ X_0(2900)$ at the quark level. \label{Fig:Quarklevel} }
\end{figure}

\subsection{Production Mechanism}

For the initial step, we analyze the process $B^+ \to D^+ X_0(2900)$ at the quark level, as shown in Fig.~\ref{Fig:Quarklevel}. The anti-bottom quark transits into an anti-charm quark by emitting a $W^+$ boson with the up quark as a spectator. Subsequently, the $W^+$ boson couples to $c\bar{s}$ quark pair, this weak transition process is phenomenologically described by the operator $H_W$. Following with the creation of $d\bar{d}$ from the vacuum, the $c$ quark and the $\bar{d}$ quark hadronize into a $D^+$ meson, while the remaining quarks hadronize into the $X_0(2900)$. To capture this hadronization process, we employ an operator $H_T$. Therefore, the process $B^+ \to D^+ X_0 (2900)$ can be described by the product of $\mathcal{H}_W$ and $\mathcal{H}_T$, denoted as, $\mathcal{H}=\mathcal{H}_W \mathcal{H}_T$. The direct estimations of the diagram in Fig.~\ref{Fig:Quarklevel} are rather difficult. Therefore, in the present work, we simply the estimations by incorporating a complete basis composed of two mesons. As a result, the decay amplitude becomes, 
\begin{eqnarray}
	&&\langle D^+ X_0(2900) \left|  \mathcal{H}\right| B^+ \rangle \nonumber \\&& \qquad =\sum_{M_1M_2} \langle D^+ X_0(2900) \left|  \mathcal{H}_T \left| M_1 M_2 \rangle \langle M_1 M_2 \right| \mathcal{H}_W \right| B^+ \rangle. \quad
\end{eqnarray}  

\begin{figure}[t]
	\centering
	\begin{tabular}{cc}\includegraphics[scale=0.4]{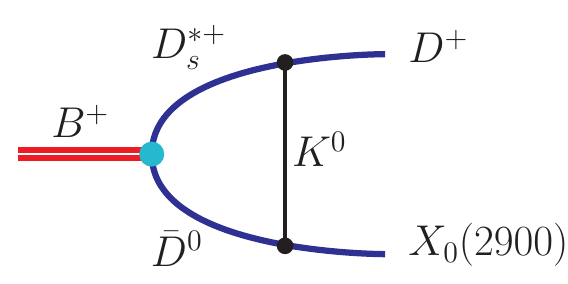}&
	\includegraphics[scale=0.4]{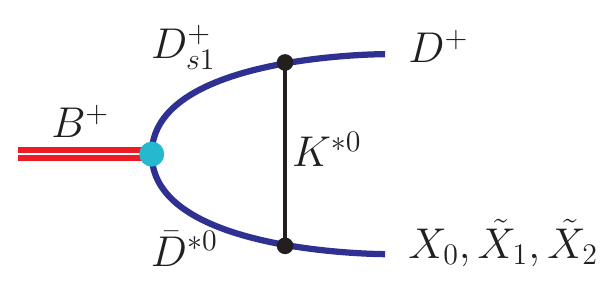}\\
$(a)$ &  $(b)$
	\end{tabular}	
    \caption{(Color online.) Diagram contributing to $B^+\to D^+ X_0(2900)$ and $B^+ \to D^+ \tilde{X}_{1,2}$ at the hadron level. Diagram (a) is the one considered in Ref.~\cite{Chen:2020eyu}, while diagram (b) will be considered in the present estimations.  \label{Fig:Hadronlevel} }
\end{figure}

In principle, all the possible  $|M_1M_2\rangle$ meson pair that could connect the initial $B^+$ and final $D^+ X_0(2900)$ states should be considered. However, in practice, only the dominant contributions are taken into account at the hadron level. For instance, in Ref.~\cite{Chen:2020eyu}, the authors investigate the process $B^+ \to D^+ X_0(2900)$ by assuming the loop in Fig.~\ref{Fig:Hadronlevel}-(a) as the dominant contributions. This choice is based on the observation that the branching fraction of $B^+\to D_s^{\ast+} \bar{ D}^0$ is $(7.6\pm 1.6)\times 10^{-3}$, which is approximately two order of magnitude larger than that of $B^+ \to D^+ X_0(2900)$, and $X_0(2900)$ predominantly decay into $\bar{D}K$ final states\footnote{In Ref.~\cite{Chen:2020eyu}, the authors investigated the process $B^- \to D^- \bar{X}_0(2900)$. The mechanism operating in $B^- \to D^- \bar{X}_0(2900)$ should be the same as $B^+ \to D^+ X_0(2900)$.}. However, in Ref.~\cite{Chen:2020eyu}, when the authors estimated the branching fractions of $B^- \to D^- X_0(2900)$ from the fit fraction of $X_0(2900)$, a factor 2 resulting from the branching fractions of $X_0(2900)\to D^+ K^-$ being $50\%$ is missed, then, the obtained the branching fraction of $B^+ \to D^- X_0(2900)$ is underestimated with a factor $1/2$. For the same reason, the coupling constant $g_{X_0 D^0 \bar{K}^0}$ is overestimated with a factor $2$. Thus, in their claimed parameter range, the branching fraction of $B^- \to D^- X_0(2900)$ contributing from diagram (a) is only $1/4$ of the measured one, which indicates that the mechanism as shown in diagram (a) should not be the dominant one in the process $B^+ \to D^+ X_0(2900)$.

\begin{figure}[t]
	\centering
	\begin{tabular}{cc}\includegraphics[scale=0.4]{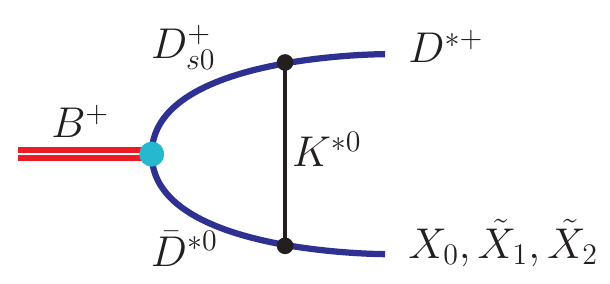}&
		\includegraphics[scale=0.4]{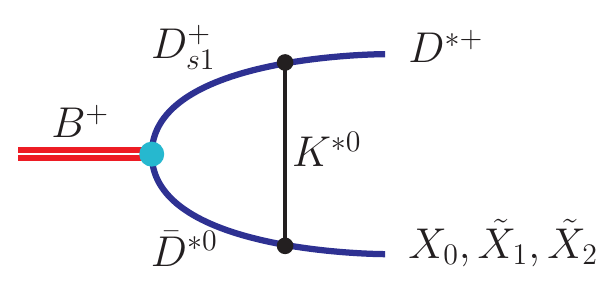}\\
		$(a)$ &  $(b)$
	\end{tabular}	
	\caption{(Color online.) Diagram contributing to $B^+\to D^{*+} X_0,\tilde{X}_{1,2}$ at the hadron level. \label{Fig:HadronLevel2} }
\end{figure}

In the molecular frame, $X_0(2900)$ is composed of $\bar{D}^\ast$ and $K^\ast$, indicating a strong coupling between $X_0(2900)$ and its components $\bar{D}^\ast K^\ast$. Additionally, the branching fractions of $B^+ \to D_{s1}^+(2460) \bar{D}^\ast$ is measured to be $(1.2\pm 0.3)\%$, which is approximately three orders of magnitude larger than that of $B^+ \to D^+ X_0(2900)$. Moreover, the charmed-strange meson $D_{s1}(2460)$ couples to $D^+ K^\ast$ via $S$-wave. Thus, we anticipate that the digram depicted in Fig.~\ref{Fig:Hadronlevel}-(b) should play a crucial role in the process $B^+\to D^+ X_0(2900)$ if $X_0(2900)$ is assigned as a $\bar{D}^\ast K^\ast $ molecular state. Similarly, the spin partner of $X_0(2900)$, $\tilde{X}_1$ and $\tilde{X}_2$, could be reproduce with the same mechanism. In addition, if one replace the final $D^+$ with $D^{\ast+}$, the intermediate states could be both $D_{s1}^+$ and $D_{s0}^{+}$, since both $D_{s0}^+$ and $D_{s1}^+$ couple to $D^{\ast+} K^{\ast 0}$ via $S$-wave. The diagrams contributing to $B^+\to D^{\ast+} X_0, \tilde{X}_{1,2}$ are presented in Fig.~\ref{Fig:HadronLevel2}.

 In our current study, we evaluate the contributions of Fig.~\ref{Fig:Hadronlevel}-(b) to $B^+\to D^+ X_0(2900)$ and compare our estimations with the experimental data in Eq.~(\ref{Eq:BRexp}), and this comparison could be a crucial test to the $\bar{D}^\ast K^\ast $ molecular interpretations to the $X_0(2900)$. In addition, with the parameter range determined by the process $B^+ \to D^+ X_0(2900)$, we can evaluate the branching fractions of $B^+\to D^+ \tilde{X}_{1,2}$ and $B^+ \to D^{\ast +} X_0,\tilde{X}_{1,2}$, which may provide some helpful information for searching the spin partner of $X_0(2900)$ in the $B^+$ decay processes.

\subsection{Effective Lagrangian}
As discussed above, the contribution of the loop illustrated in Fig.~\ref{Fig:Hadronlevel}-(b) and Fig.~\ref{Fig:HadronLevel2} are anticipated to be  dominant for the process $B^+ \to D^{(\ast)+} X_0, \tilde{X}_1, \tilde{X}_2$ when considering $X_0(2900)$ as the $\bar{D}^\ast K^\ast$ molecular state with $I(J^P)=0(0^+)$, while $\tilde{X}_1, \tilde{X}_2$ are the spin partner of $X_0(2900)$ with $J=1$ and $J=2$, respectively. In this study, all the hadron interactions are described by the effective Lagrangian. As for the effective Lagrangian for the coupling between $X_0$ and $\tilde{X}_{1,2}$ with its component, we employ the simple $S$-wave coupling, which are~\cite{Xiao:2020ltm},
\begin{eqnarray}
	\mathcal{L}_{X_0D^*K^*} &=& g_{X_0}X_0 \left[{D}^{*-\mu} K^{*+}_\mu-\bar{D}^{*0\mu} K^{*0}_\mu \right] ,\nonumber\\
	\mathcal{L}_{\tilde{X}_1D^*K^*} &=& g_{\tilde{X}_1}\varepsilon_{\mu\nu\alpha\beta}\partial^\mu \tilde{X}_1^\nu \left[{D}^{*-\alpha} K^{*+\beta}-\bar{D}^{*0\alpha} K^{*0\beta} \right],\\
	\mathcal{L}_{\tilde{X}_2D^*K^*} &=& g_{\tilde{X}_2}\tilde{X}_2^{\mu\nu} \left[{D}^{*-}_
	\mu K^{*+}_\nu-\bar{D}^{*0}_\mu K^{*0}_\nu \right] ,\nonumber
\end{eqnarray}
where $g_{X_0}$ and $g_{\tilde{X_1}/\tilde{X_2}}$ are the effective coupling constants, and their values will be discussed later. As for the weak interaction vertex $B^+ \to D_{s1}^+(2460)/D_{s0}^+(2317) \bar{D}^{\ast 0}$, we utilize the parametrized hadronic matrix elements obtained by the effective Hamiltonian at the quark level, which are~\cite{Cheng:2003sm,Soni:2021fky},
   \begin{eqnarray}
&&\langle0|J_\mu|D_{s0}(p_1)\rangle = -i f_{D_{s0}} p_{1\mu} \;,\nonumber\\   
&&\langle0|J_\mu|D_{s1}(p_1,\epsilon)\rangle =f_{D_{s1}} \epsilon(p_1)_\mu m_{D_{s1}} \;,\nonumber\\
&&\langle D^\ast (p_2,\epsilon)|J_\mu|B(p_0)\rangle\nonumber\\ 
&&\qquad=\frac{i\epsilon^\nu}{m_{B}+m_{D^\ast}}\Bigg\{i\varepsilon_{\mu\nu\alpha\beta}P^\alpha q^\beta V(q^2)\nonumber \\&&\qquad -\left(m_{B}+m_{D^\ast}\right)^2 g_{\mu\nu}A_1(q^2) +P_\mu P_\nu A_2(q^2)\nonumber\\
&&\qquad +2m_{D^\ast} \left(m_{B}+m_{D^\ast}\right)\frac{P_\nu q_\mu}{q^2}\left[A_3(q^2)-A_0(q^2)\right]\Bigg\},\qquad \qquad \;
   \end{eqnarray}
 with $J_\mu = \bar{q}_1 \gamma_\mu(1-\gamma_5)q_2$, $P_\mu=p_{0\mu} +p_{2\mu}$ and $q_\mu =p_{0\mu}-p_{2\mu}$. $A_{0,1,2}(q^2)$ and $V(q^2)$ are the weak transition form factors, while $A_3(q^2)$ is the linear combination of form factors $A_1(q^2)$ and $A_2(q^2)$, which is~\cite{Cheng:2003sm},
 \begin{equation}
A_3(q^2)=\frac{m_B+m_{D^\ast}}{2m_{D^\ast}}A_1(q^2)-\frac{m_B-m_{D^\ast}}{2m_{D^\ast}}A_2(q^2).
 \end{equation}
 
With the above hadronic matrix elements, we can get the amplitude of weak interaction vertex $B^+ \to D_{sJ}^+\bar{D}^*$, which are,
\begin{eqnarray}
&&\mathcal{A}(B^+ \to D_{sJ}^+ \bar{D}^*) \nonumber\\
	&& \qquad = \frac{G_F}{\sqrt{2}}V_{cb}V_{cs}a_1	\left \langle D_{sJ}^+ \left | J_\mu \right |0  \right \rangle 	\left \langle \bar{D}^{*0} \left | J^\mu \right |B^+  \right \rangle,
\end{eqnarray}
where $ G_F$ is the Fermi constant, $V_{bc}$ and $V_{cs}$ are the CKM matrix elements. $a_1 = c_1^{eff} + c_2^{eff}/N_c$ with $c_{1,2}^{eff}$ being the effective Wilson coefficients, obtained by the factorization approach~\cite{Bauer:1986bm}, and $N_c=3$ is the number of the quark color. In the present work, we adopt $ G_F=1.166\times10^{-5}\,{\rm GeV^{-2}}$, $V_{cb}=0.041$, $V_{cs}=0.987$ and $a_1=1.05$, $f_{D_{s1}}=158~{\rm MeV}$, $f_{D_{s0}}=70~{\rm MeV}$ as in Refs.~\cite{ParticleDataGroup:2020ssz,Ali:1998eb,Ivanov:2006ni,Hwang:2004kga}.

As for the coupling relevant to the charmed mesons, they can be constructed in the heavy quark limit and chiral symmetry. In the heavy quark limit, the $S-$wave heavy-light mesons can be described by the superfield $H^{Q}_a$, which is
\begin{eqnarray}
	H_a^{(Q)}=\frac{1+\slashed{v}}{2}[P_a^{*(Q)\mu}\gamma_\mu-P_a^{(Q)}\gamma_5], 
\end{eqnarray}
where $P^{\ast}$ and $P$ refer to the vector and pseudosclar $S$-wave heavy light mesons, respectively.

As for the $P-$wave heavy-light mesons, they are devided into two doublets depending on the light degree of freedom, which are $S-$double with $s_\ell=1/2$ and $T-$doublet with $s_\ell=3/2$. The involved $D_{s1}(2460)$ and $D_{s0}(2317)$ belong to the $S-$doublet, and the concrete form of the $S-$double superfield is,  
\begin{eqnarray}	 
	&&S_a^{(Q)}=\frac{1+\slashed{v}}{2}[P_{1a}^{\prime(Q)\mu}\gamma_\mu\gamma_5-P_{0a}^{*(Q)}],
\end{eqnarray}
with $P_1^\prime$ and $P_0^{\ast}$ to be $P-$wave heavy light meson in the $S$-doublet with $J=1$ and $J=0$, respectively.   

In terms of heavy quark limit and chiral symmetry, the effective Lagrangians has been constructed in the literatures as~ \cite{Casalbuoni:1996pg,Casalbuoni:1992gi,Casalbuoni:1992dx},
\begin{eqnarray}\label{eq:LagrangianV}
	\mathcal{L}_{V}&=&i\zeta\left\langle H_{b}^{(Q)} \gamma^{\mu}\left(-\rho_{\mu}\right)_{b a} \bar{S}_{a}^{(Q)}\right\rangle \nonumber \\
	&&+i \mu\left\langle H_{b}^{(Q)} \sigma^{\lambda \nu} F_{\lambda \nu}(\rho)_{b a} \bar{S}_{a}^{(Q)}\right\rangle +\text { H.c,}
\end{eqnarray}
where $F_{\mu\nu}(\rho)=\partial_{\mu}\rho_{\nu}-\partial_{\nu}\rho_{\mu}+[\rho_{\mu},\rho_{\nu}]$, and $\rho_{\mu}$ is defined as
\begin{eqnarray}
	\rho_{\mu}=i\frac{g_V}{\sqrt{2}}\mathcal{V}_{\mu}.
\end{eqnarray}
$\mathcal{V}_{\mu}$ is  matrix form of the light vector mesons with the concrete form as, 
\begin{eqnarray}\label{eq:matrix V}
	\mathcal{V} &=& \left(\begin{array}{ccc}\frac{\rho^0} {\sqrt {2}}+\frac {\omega} {\sqrt {2}}&\rho^+ & K^{*+} \\
		\rho^- & -\frac {\rho^0} {\sqrt {2}} + \frac {\omega} {\sqrt {2}} & K^{*0} \\
		K^{*-}& {\bar K}^{*0} & \phi \\
	\end{array}\right).
\end{eqnarray}

Since we only focus on the S-wave coupling in the present work, which means the term with constant $\mu$ in Eq.~\eqref{eq:LagrangianV} should be neglected, so we obtained the detailed Lagrangian of $D_{1}^\prime D^{(*)}K^*$ and $D_0 D^{\ast }K^\ast$, which reads,
\begin{eqnarray}\label{Eq:LD1DV}
	&&\mathcal{L}_{\mathcal{D}_1^\prime \mathcal{D}\mathcal{V}} = g_{\mathcal{D}_1^\prime\mathcal{D}\mathcal{V}}\mathcal{D}_{1b}^{\prime \mu} \mathcal{V}_{\mu_{ba}}\mathcal{D}_a^\dagger \nonumber \\
	&&\mathcal{L}_{\mathcal{D}_1^\prime \mathcal{D}^\ast\mathcal{V}} = i~ g_{\mathcal{D}_1^\prime \mathcal{D}^\ast\mathcal{V}}\varepsilon_{\mu\nu\alpha\beta}\mathcal{V}_{ba}^{\ast\nu}\mathcal{D}_b^{\ast\mu}\overleftrightarrow{\partial^\beta} \mathcal{D}_{1a}^{\prime\alpha\dagger} \\
	&&\mathcal{L}_{\mathcal{D}_0^\ast  \mathcal{D}^\ast \mathcal{V}} = g_{\mathcal{D}_0^\ast \mathcal{D}^\ast \mathcal{V}}\mathcal{D}_{b}^\mu \mathcal{V}_{\mu_{ba}}\mathcal{D}_{0a}^{\ast \dagger} \nonumber
\end{eqnarray} 
where the $\mathcal{D}$ is charmed meson triplets $(D^0, D^+, D_s^+)$, while $D^\ast$, $D_1^\prime$, and $D_0^\ast$ have the similar definitions. The relevant coupling constants will be discussed later.

\subsection{Decay Amplitude}
With the above effective Lagrangians, the amplitude of $B^{+}(p_0)\to[ D_{s1}^{\prime +}(p_1) \bar{D}^0(p_2)] K^{\ast 0} (q)\to D^+(p_3) X_0(2900)(p_4) $ corresponding to Fig.~\ref{Fig:Hadronlevel}-(b) could be obtained as,
\begin{eqnarray}	
\mathcal{M} &=& i^3\int \frac{d^4 q}{(2\pi)^4} \left[\mathcal{A}_{\mu \nu } ^{B\to D_{s1}^\prime \bar{D}^*} (p_1,p_2)\right]\left [ i g_{D_{1}^\prime DV}  g_{\alpha \beta}\right ]  
	\left [ i g_{X_0} g_{\rho\phi} \right ]\nonumber\\
	&&\times \left [ \frac{-g^{\mu \alpha} +p_1^\mu p_1^\alpha / m_1^2}{p_1^2-m_1^2}  \right]
	\left [ \frac{-g^{\nu \rho} +p_2^\nu p_2^\rho / m_2^2}{p_2^2-m_2^2}  \right ] 
	 \nonumber\\&&\times \left [ \frac{-g^{\beta \phi} +q^\beta q^\phi / m_q^2}{q^2-m_q^2}  \right ] \mathcal{F}^2\left ( q^2, m_q^2\right ). 
\end{eqnarray}
The rest amplitudes corresponding to Fig.~\ref{Fig:Hadronlevel}-(b) and Fig.~\ref{Fig:HadronLevel2} can be found in Appendix.~\ref{Sec:AppendixA}.

In the above amplitude, we introduce a form factor $\mathcal{F}(q^2,m^2)$ in monopole form to compensate for the off-shell effect and to avoid the ultraviolet divergence in the loop integral. Its concrete form is,
\begin{equation}
	\mathcal{F}(q^2,m^2) = \frac{m^2-\Lambda^2}{q^2-\Lambda^2},
\end{equation}
where ${\rm \Lambda} = m + \alpha {\rm \Lambda_{QCD}}$ is the cut-off parameter~\cite{Cheng:2004ru}, with ${\rm \Lambda_{QCD}}$ = 220 MeV and $\alpha$ to be the model parameter. It should be noted that the form factor also plays the role of describing the momentum dependence of the coupling between $X_0(2900)$ and its component as indicated in Ref.~\cite{Gamermann:2009uq,Bayar:2023azy}.

Then, the partial width of $B^{+} \to D^+ X_0(2900) $ can be obtained by,
\begin{equation}
	\Gamma_{B^+ \to D^+ X_0} = \frac{1}{8\pi} \frac{|\vec{p}\,|}{m_{B}^2}\overline{\left|\mathcal{M}_{B^{+} \to D^+ X_0 }\right|^2},
\end{equation}
and consequently, we can obtain the branching fraction of $B^+                           \to D^+ X_0(2900)$ depending on the model parameter $\alpha$.

\section{Numerical Results and Discussions}
\label{Sec:Num}
\subsection{Coupling Constants}
 
	Before we estimate the branching ratio of $B^+ \to D^{(*)+} X(2900)$, some relevant coupling constants should be further clarified. The coupling constant of $X(2900), \tilde{X}_{1,2}$ with their components could be estimated by compositeness conditions in the molecular frame. In Ref.~\cite{Xiao:2020ltm}, the author estimated the decay properties of $X_0$ and $\tilde{X}_{1,2}$ in the molecular frame and found the measured width of $X_0(2900)$ could be well reproduced. In the considered model parameter range, the coupling constants $g_{X_0}$ and $g_{\tilde{X}_{1,2}}$ was estimated to be approximately 9.11 GeV, 3.84 and 16.28 GeV, respectively.   
 
As for the weak transition form factor, the estimations in Refs.~\cite{Cheng:2003sm,Soni:2021fky} parametrized them in the form,
\begin{equation}\label{eq:FFab}
	F(Q^2) = \frac{F(0)}{1-a\zeta+b\zeta^2}
\end{equation}
with $\zeta = Q^2/m_{B}^2$. The relevant parameters $F(0)$, $a$, and $b$ for each form factor are collected in Table~\ref{Tab:F0ab}. In order to simplify the estimations, we further parametrize Eq.~\eqref{eq:FFab} in the form,
\begin{equation}\label{eq:FFL1L2}
	F(Q^2) = F(0)\frac{\Lambda_1^2}{Q^2-\Lambda_1^2}\frac{\Lambda_2^2}{Q^2-\Lambda_2^2}.
\end{equation}
By fitting Eq.~\eqref{eq:FFab} with Eq.~\eqref{eq:FFL1L2}, we can obtain the values of $\Lambda_1$ and $\Lambda_2$ for each form factor, which are listed in Table~\ref{Tab:L1L2}. 

Considering the heavy quark limit and chiral symmetry, the coupling constants of $D_{1,0}D^{(*)}V$ in Eq.~\eqref{Eq:LD1DV} are given as \cite{Ding:2008gr},
\begin{eqnarray}
	&&g_{D_1DV} = -\sqrt{2} \zeta g_V \sqrt{M_D M_{D_1}} \nonumber \\
	&&g_{D_0D^*V} = -\sqrt{2} \zeta g_V \sqrt{M_{D^*} M_{D_0}} \\
	&&g_{D_1D^*V} = \sqrt{2} \zeta g_V \nonumber,  
\end{eqnarray}
with $g_V \simeq 5.8$ by imposing the Kawarabayashi-Suzuki-Riazuddin-Fayyazuddin relations~\cite{Ding:2008gr} and $\zeta = 0.1$~\cite{Casalbuoni:1996pg}.

\begin{table}
  \caption{\label{Tab:F0ab}The values of the parameters $F(0)$, $a$ and $b$ in the form factors~\cite{Verma:2011yw}.}
\renewcommand\arraystretch{1.5}
	\begin{tabular}{p{1.9cm}<\centering p{1.9cm}<\centering p{1.9cm}<\centering p{1.9cm}<\centering}
		\toprule[1pt]
		Parameter &$F(0)$ &$a$ &$b$ \\
		\midrule[1pt]
		$A_0$ &0.68 &1.21 &0.36 \\
		$A_1$ &0.65 &0.60 &0.00 \\
		$A_2$ &0.61 &1.12 &0.31 \\
		$V$ &0.77 &1.25 &0.38 \\
		\bottomrule[1pt]	 
	\end{tabular}
\end{table}

\begin{table}
\caption{\label{Tab:L1L2}Values of the parameters $\Lambda_1$ and $\Lambda_2$ obtained by fitting Eq.~\eqref{eq:FFab} with Eq.~\eqref{eq:FFL1L2}.}
\renewcommand\arraystretch{1.5}
	\begin{tabular}{p{2cm}<\centering p{1.4cm}<\centering p{1.4cm}<\centering p{1.4cm}<\centering p{1.4cm}<\centering}
		\toprule[1pt]
		Parameter &$A_{0}$ &$A_{1}$ &$A_{2}$ &$V$ \\
		\midrule[0.8pt]
		$\Lambda_1$ &6.5 &8.95 &6.65 &6.30 \\
	    $\Lambda_2$ &7.1 &8.75 &7.55 &7.10 \\
        \bottomrule[1pt]
	\end{tabular}
\end{table}

\begin{figure}[htb]
	\centering
	 \vspace{0.5cm}
	\includegraphics[scale=0.8]{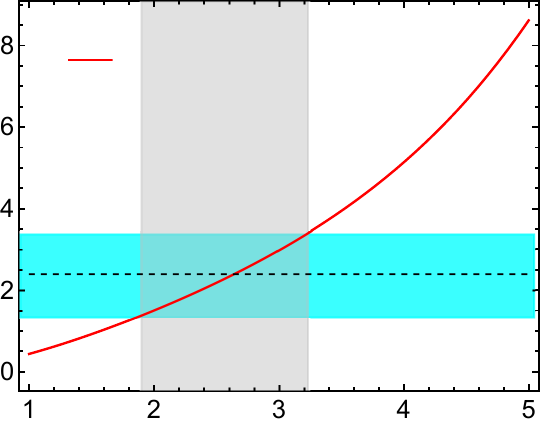}
	\put(-102,-10){\large  $\alpha$} 
	\put(-220,35){\rotatebox{90}{\text { \footnotesize Branching ratio $(\times10^{-5})$ }}}%
        \put(-160,136){$B^+\to D^{+} X_0$}
	\caption{(Color online.) The branching ratio of $B^{+} \to X_0(2900) D^+$ (in unit of $10^{-5}$) depending on the model parameter $\alpha$. The horizontal black dashed line with cyan band indicates the measured branching fraction of $B^{+} \to X_0(2900) D^+$, while the vertical light grey band refers to the $\alpha$ range in which the measured branching fraction could be reproduce. \label{Fig:BR-DX0}}
\end{figure}

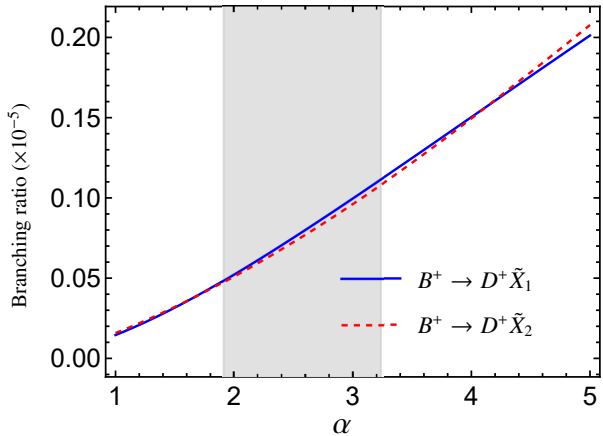
\begin{figure}[htb]
	\centering
	\vspace{0.5cm}
	\includegraphics[scale=0.8]{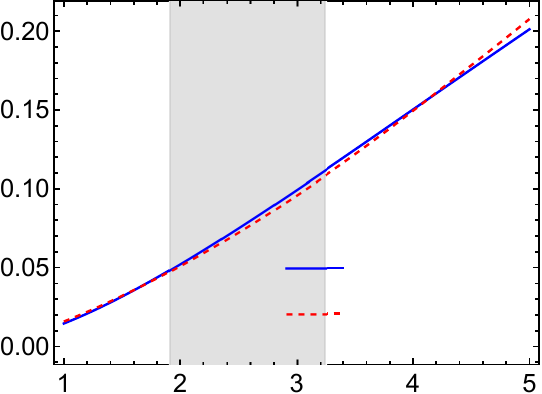}
	\put(-102,-10){\large  $\alpha$} 
	\put(-225,35){\rotatebox{90}{\text { \footnotesize Branching ratio $(\times10^{-5})$ }}}%
        \put(-70,45){$B^+\to D^{+} \tilde{X}_1$}
        \put(-70,28){$B^+\to D^{+} \tilde{X}_2$}
	\caption{(Color online.) The branching fraction of $B^+ \to D^+ \tilde{X}_{1/2}(2900)$ (in unit of $10^{-5}$) depending on the model parameter $\alpha$. The vertical light grey band refers to the $\alpha$ range determined by $B^+ \to D^+ X_0(2900)$.
 \label{Fig:BR-DX1X2}}
\end{figure}

\subsection{Branching Fraction}

With above preparations, all the relevant coupling constants have been fixed except the model parameter $\alpha$. This phenomenological model parameter cannot be determined by first principle methods. As indicated in Ref.~\cite{Cheng:2004ru}, $\alpha$ should be of order unity. In the present estimation, we  vary $\alpha $ from $1$ to $5$ to check the parameter dependence of the branching fraction. By reproducing the experimental data in Eq.~\eqref{Eq:BRexp}, we can determine the value of parameter $\alpha$ and check the rationality of the parameter range, which can further test the $\bar{D}^\ast K^\ast $ molecular interpretations to the $X_0(2900)$. In addition, consider the similarity of $X_0(2900)$, $\tilde{X}_1$ and $\tilde{X}_2$, we can evaluate the production ratios of $B^+ \to D^+ \tilde{X}_{1,2}$ and $B^+ \to D^{\ast +} X_0,\tilde{X}_{1,2}$ in the same model parameter range, which should be helpful for experimentally searching for the spin partner of $X_0(2900)$.

The branching ratio of $B^+ \to D^+ X_0(2900)$ depending on the model parameter $\alpha$ is presented in Fig.~\ref{Fig:BR-DX0}, where the experimental measurement from the LHCb Collaboration is also presented for comparison. From the figure, one can find that the branching fraction increases with the increasing of model parameter $\alpha$. Our estimation could well reproduce the experimental data in the range $1.9<\alpha <3.2$, which is of order unity. Thus, we can conclude that the $\bar{D}^\ast K^\ast $ molecular interpretations to the $X_0(2900)$ should be reasonable. 

In the similar way, we can investigate the process $B^+ \to D^+ \tilde{X}_{1,2}$, where $\tilde{X}_{1}$ and $\tilde{X}_{2}$ are the spin partner of $X_0(2900)$ with $J=1$ and $J=2$, respectively. The branching ratios of $B^+ \to D^+ \tilde{X}_{1,2}$ depending on the model parameter $\alpha$ are presented in Fig.~\ref{Fig:BR-DX1X2}. Our estimations indicate that these two branching ratios are very close but one order of magnitude smaller than that of $B^+ \to D^+ X_0(2900)$. In particular, in the parameter range determined by $B^+ \to D^+ X_0(2900)$, the branching ratios of $B^+ \to D^+ \tilde{X}_{1,2}$  are predicted to be $(0.5\sim 1.1) \times 10^{-6}$. As indicated in Ref.~\cite{Xiao:2020ltm}, $\tilde{X}_{1,2}$ predominantly decay into $\bar{D}^\ast K$, and the branching ratios of $\tilde{X}_{1,2}\to \bar{D}^{\ast-} K^+$ are about $40\%$. Thus, we can approximately obtain the branching fraction of $B^+ \to D^+ \tilde{X}_{1,2}\to D^+ D^{\ast-} K^+ $ to be of order of $10^{-7}$, which indicates the fit fractions of $\tilde{X}_{1,2}$ in $B^+ \to D^+ D^{\ast-} K^+ $ is of the order $10^{-3}$. Thus, it's impossible to search the spin partner of $X_0(2900)$ in the process $B^+ \to D^+ D^{\ast-} K^+ $.

\begin{figure}[t]
	\centering
	\vspace{0.5cm}	\includegraphics[scale=0.60]{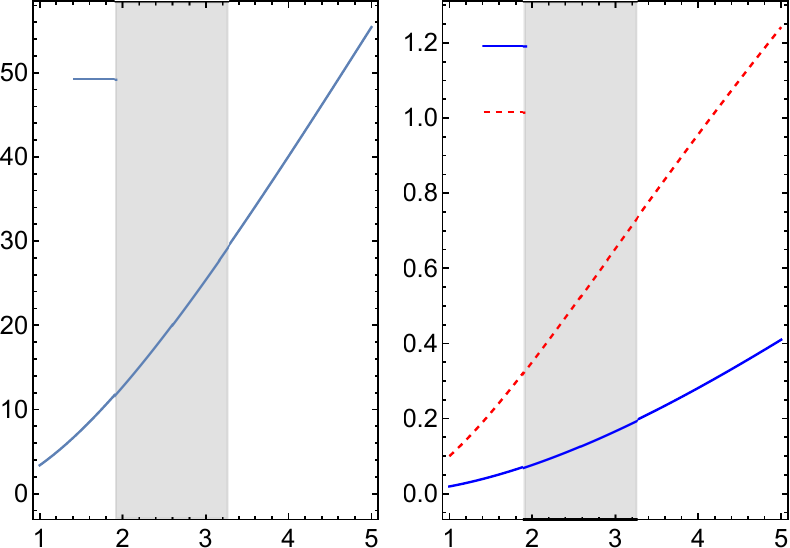}
	\put(-110,-10){\large $\alpha$} 
	\put(-240,35){\rotatebox{90}{\text { \footnotesize Branching ratio $(\times10^{-5})$ }}}%
        \put(-188,133){$B^+\to D^{*+}\tilde{X}_1$}
        \put(-72,143){$B^+\to D^{*+} X_0$}
        \put(-72,123){$B^+\to D^{*+}\tilde{X}_2$}
	\caption{(Color online.) The branching fractions of $B^+ \to D^{\ast+} X_0(2900), \tilde{X}_{1,2}$ (in unit of $10^{-5}$) depending on the model parameter $\alpha$. The vertical light grey band refers to the $\alpha$ range determined by $B^+ \to D^+ X_0(2900)$. \label{Fig:BR-DxX0X1X2}}
\end{figure}

In addition to $B^+ \to D^+ X_0(2900), \tilde{X}_{1,2}$, we also investigate the production process $B^+ \to D^{\ast +} X_0(2900),\tilde{X}_{1,2}$ in the present work. The branching fractions of the relevant processes depending on the model parameter $\alpha$ are presented in Fig.~\ref{Fig:BR-DxX0X1X2}. In the left panel, we present the branching fractions of the process $B^+ \to D^{\ast +} \tilde{X}_1$. The present estimations indicate that in the parameter range determined by $B^+\to D^+ X_0(2900)$, the branching fraction of $B^+ \to D^{\ast +} \tilde{X}_1$ is $(1.2\sim 2.9) \times 10^{-4}$, which is about one order of magnetite larger than that of $B^+ \to D^+ X_0(2900)$. As listed in Table~\ref{Tab:BR}, the branching fraction of $B^+ \to D^{\ast +} D^{\ast -} K^+$ is $(1.32 \pm 0.18) \times 10^{-3}$, thus one can approximately obtain the fit fraction of $B^+\to D^{*+} \tilde{X}_1 \to D^{*+}D^{*-}K^+$ is about $6\%$ with the assumption that  the branching fraction of $\tilde{X}_1 \to D^{\ast-} K^+$ is about $40\%$ ~\cite{Xiao:2020ltm}. Comparing to the fit fraction of $X_0(2900)$ in the process $B^+ \to D^+ D^- K^+$ being $(5.6\pm 1.4 \pm 0.5)\%$, we can conclude that experimentally searching for $\tilde{X}_1 $ in the process $B^+ \to D^{\ast +} D^{\ast-} K^+$ is possible. The predicted branching fractions of the processes $B^+ \to D^{\ast +} X_0(2900)$ and $B^+ \to D^{\ast +} \tilde{X}_2$ are present in the right panel of Fig.~\ref{Fig:BR-DxX0X1X2}. In the determined $\alpha$ range, the branching fraction of $B^+ \to D^{\ast +} X_0(2900)$ is estimated to be  $(0.7 \sim 1.9) \times 10^{-6}$, which is at least one order smaller than that of $B^+ \to D^+ X_0(2900)$. As for $B^+ \to D^{\ast +} \tilde{X}_2$, the predicted branching fraction is $(3.2\sim 7.2) \times 10^{-6}$, which is about several times smaller than that of $B^+ \to D^+ X_0(2900)$. Considering the branching fraction of $B^+ \to D^{\ast + } D^{\ast -} K^+$ is of the order $10^{-3}$, the fit fraction of $\tilde{X}_2$ in $B^+ \to D^{\ast + } D^{\ast -} K^+$ is of the order $10^{-3}$, which indicate that searching for $X_0(2900)$ in the process $B^+ \to D^{\ast +} D^- K^+$  and searching for $\tilde{X}_2$ in the process $B^+ \to D^{\ast +} D^{\ast-} K^+$ are both impossible. In Table~\ref{Tab:Prediction}, we summarized the predicted branching fractions of the relevant processes and the experimental potential of searching for $X_0(2900)$ and its spin partner in the corresponding processes. 

\begin{table}[t]    \centering\caption{Branching ratios of all decay modes considered in the present work. The symbols \Checkmark and \XSolidBrush indicate the experimental potential of searching $X_0(2900)$ and its spin partner in the corresponding processes. \label{Tab:Prediction}}
\renewcommand\arraystretch{1.5}
    \begin{tabular}{p{2.0cm}<\centering p{4.2cm}<\centering p{1.5cm}<\centering }
    \toprule[1pt]
       Process  &  Branching ratio & Status \\
    \midrule[1pt]
        $B^+ \to D^+X_0$ & $(1.4\sim2.4) \times 10^{-5}$ & Input \\
        $B^+ \to D^+\tilde{X}_1$ & $(0.5\sim1.1) \times 10^{-6}$ & \XSolidBrush \\
        $B^+ \to D^+\tilde{X}_2$ & $(0.5\sim1.1) \times 10^{-6}$& \XSolidBrush \\
        $B^+ \to D^{*+}X_0$ & $(0.7\sim1.9) \times 10^{-6}$& \XSolidBrush \\
        $B^+ \to D^{*+}\tilde{X}_1$ & $(1.2\sim 2.9) \times 10^{-4}$&\Checkmark \\
        $B^+ \to D^{*+}\tilde{X}_2$ & $(3.2\sim 7.2) \times 10^{-6}$ & \XSolidBrush\\
    \bottomrule[1pt]
    \end{tabular}
    \label{tab:my_label}
\end{table}

\section{SUMMARY}
Since the observation of $X_{0,1}(2900)$ by the LHCb collaboration, various interpretations have been proposed to investigate their internal structure. In particular, the tetraquark interpretations were proposed inspired by the fully open flavor properties of $X_{0,1}(2900)$. In addition, the observed mass of $X_{0,1}(2900)$ is close the threshold of $\bar{D}^\ast K^\ast$, and the $J^P$ quantum numbers of the $S-$wave $\bar{D}^\ast K^\ast$ molecule could be consistent with those of $X_0(2900)$, thus, the interpretation have been proposed for the $X_0(2900)$. The mass spectroscopy and decay behavior of $X_0(2900)$ have been reproduced in the molecular frame. To further test the rationality of the molecular interpretation, we investigate the $X_0(2900)$ production process $B^+ \to D^+ X_0(2900)$  in present work. 

In the molecular frame, $X_0(2900)$ is composed of $\bar{D}^\ast$ and $K^\ast$, and we also find that the branching fraction of $B^+ \to D_{s1}^+(2460) \bar{D}^{\ast 0}$ is measured to be order of $10^{-2}$, which is approximately three orders of magnitude larger than that of $B^+ \to D^+ X_0(2900)$. Additionally,  it is worth noting that $D_{s1}^+(2460)$ couples to $D^+ K^\ast$ through an $S$-wave interaction. Thus, we can infer that the initial $B^+$ meson and the final $D^+ X_0(2900)$ might be connected through $D_{s1}^+(2460) \bar{D}^{\ast 0} $ meson pair by exchanging a $K^{\ast 0}$ meson. Similarly, the spin partner of $X_0(2900)$ could be produced in the process $B^+ \to D^+ \tilde{X}_{1,2}$. In addition, the productions of  $X_0(2900)$ and its spin partners in the processes $B^+ \to D^{\ast +} X_0(2900), \tilde{X}_{1,2}$ are also investigated in the present work. 

Our estimations indicate that the branching fractions of $B^+ \to D^+ X_0(2900)$ could be well reproduced in the parameter range $1.9 <\alpha <3.2$,  which indicate that the $\bar{D}^\ast K^\ast $ molecular interpretations for the $X_0(2900)$ should be reasonable. In the same parameter range, we find the branching fractions of the processes $B^+ \to D^+ \tilde{X}_{1,2}$, $B^+ \to D^{\ast +} X_0(2900)$, and $B^+ \to D^{\ast+} \tilde{X}_2$ are of the order $10^{-6}$. While the branching fraction of the process $B^{+}\to D^{\ast +} \tilde{X}_1$ is of the order $10^{-4}$, and the fit fraction of $\tilde{X}_1$ in the process $B^+ \to D^{\ast+} D^{\ast -} K^+$ is about $6\%$, which should be accessible experimentally.

\section*{ACKNOWLEDGMENTS}
The authors would like to thank Prof. Fu-Sheng Yu for useful discussions. This work is supported by the National Natural Science Foundation of China under the Grant 11775050 and 12335001.
\begin{widetext}
\appendix
\section{DECAY AMPLITUDES}\label{Sec:AppendixA}
The amplitudes corresponding the diagrams in Fig.~\ref{Fig:Hadronlevel} and ~\ref{Fig:HadronLevel2} read, 
	\begin{eqnarray}
		\mathcal{M}_{B^+ \to D^+\tilde{X_1}} &=& i^3\int \frac{d^4 q}{(2\pi)^4} \Big[\mathcal{A}_{\mu \nu } ^{B\to D_{s1}\bar{D}^*} (p_1,p_2)\Big]\Big [ i g_{D_{1}DV}  g_{\alpha \beta}\Big]  
		\Big [ - g_{\tilde{X_1}} \varepsilon_{\mu_1\tau\rho\phi} p_4^{\mu_1}\epsilon^\tau(p_4) \Big ]\nonumber\\
		&&\times
		\left [ \frac{-g^{\mu \alpha} +p_1^\mu p_1^\alpha / m_1^2}{p_1^2-m_1^2}  \right]\left [ \frac{-g^{\nu \rho} +p_2^\nu p_2^\rho / m_2^2}{p_2^2-m_2^2}  \right ]  \left [ \frac{-g^{\beta \phi} +q^\beta q^\phi / m_q^2}{q^2-m_q^2}  \right ] \mathcal{F}^2\left ( q^2, m_q^2\right )\nonumber \\
		\mathcal{M}_{B^+ \to D^+\tilde{X_2}} &=& i^3\int \frac{d^4 q}{(2\pi)^4} \Big[\mathcal{A}_{\mu \nu } ^{B\to D_{s1}\bar{D}^*} (p_1,p_2)\Big] \Big [ i g_{D_{1}DV}  g_{\alpha \beta}\Big ]  
		\Big [ i g_{\tilde{X_2}} \epsilon_{\rho\phi}(p_4) \Big ]\nonumber\\
		&&\times \left [ \frac{-g^{\mu \alpha} +p_1^\mu p_1^\alpha / m_1^2}{p_1^2-m_1^2}  \right]
		\left [ \frac{-g^{\nu \rho} +p_2^\nu p_2^\rho / m_2^2}{p_2^2-m_2^2}  \right ]  \left [ \frac{-g^{\beta \phi} +q^\beta q^\phi / m_q^2}{q^2-m_q^2}  \right ] \mathcal{F}^2\left ( q^2, m_q^2\right )\nonumber\\
		\mathcal{M}_{B^+ \to D^{*+}X_0}^{(a)} &=& i^3\int \frac{d^4 q}{(2\pi)^4} \left[\mathcal{A}_{\nu } ^{B\to D_{s0}\bar{D}^*} (p_1,p_2)\right] \left[ i g_{D_{0}D^*V}  g_{\delta \beta} \epsilon^{\delta}(p_3) \right]  
		\left [ i g_{X_0} g_{\rho\phi}(p_4) \right]\nonumber\\
		&&\times \left [ \frac{1}{p_1^2-m_1^2}  \right]
		\left [ \frac{-g^{\nu \rho} +p_2^\nu p_2^\rho / m_2^2}{p_2^2-m_2^2}  \right ] \left [ \frac{-g^{\beta \phi} +q^\beta q^\phi / m_q^2}{q^2-m_q^2}  \right ] \mathcal{F}^2\left ( q^2, m_q^2\right )\nonumber\\
		\mathcal{M}_{B^+ \to D^{*+}X_0}^{(b)} &=& i^3\int \frac{d^4 q}{(2\pi)^4} \left[\mathcal{A}_{\mu\nu} ^{B\to D_{s1}\bar{D}^*} (p_1,p_2)\right] \left[ i g_{D_{1}D^*V}\varepsilon_{\delta\beta\alpha\beta_2}(p_1+p_3)^{\beta_2}\epsilon^{\delta}(p_3) \right]  
		\left [ i g_{X_0} g_{\rho\phi} \right] \nonumber\\
		&&\times \left [ \frac{-g^{\mu \alpha} +p_1^\mu p_1^\alpha / m_1^2}{p_1^2-m_1^2}  \right] 
		\left [ \frac{-g^{\nu \rho} +p_2^\nu p_2^\rho / m_2^2}{p_2^2-m_2^2}  \right ] \left [ \frac{-g^{\beta \phi} +q^\beta q^\phi / m_q^2}{q^2-m_q^2}  \right ] \mathcal{F}^2\left ( q^2, m_q^2\right )\nonumber\\
		\mathcal{M}_{B^+ \to D^{*+}\tilde{X}_1}^{(a)} &=& i^3\int \frac{d^4 q}{(2\pi)^4} \left[\mathcal{A}_{\nu } ^{B\to D_{s0}\bar{D}^*} (p_1,p_2)\right] \left[ i g_{D_{0}D^*V}  g_{\delta \beta} \epsilon^{\delta}(p_3) \right]  
		\left [ - g_{\tilde{X}_1} \varepsilon_{\mu_1\tau\rho\phi}p_4^{\mu_1}\epsilon^\tau(p_4) \right] \nonumber\\
		&&\times \left [ \frac{1}{p_1^2-m_1^2}  \right]
		\left [ \frac{-g^{\nu \rho} +p_2^\nu p_2^\rho / m_2^2}{p_2^2-m_2^2}  \right ]  \left [ \frac{-g^{\beta \phi} +q^\beta q^\phi / m_q^2}{q^2-m_q^2}  \right ] \mathcal{F}^2\left ( q^2, m_q^2\right )\nonumber
   \end{eqnarray}
  \begin{eqnarray}
		\mathcal{M}_{B^+ \to D^{*+}\tilde{X}_1}^{(b)} &=& i^3\int \frac{d^4 q}{(2\pi)^4} \left[\mathcal{A}_{\mu\nu } ^{B\to D_{s1}\bar{D}^*} (p_1,p_2)\right] \left[ i g_{D_{1}D^*V}  \varepsilon_{\delta\beta\alpha\beta_2}(p_1+p_3)^{\beta_2} \epsilon^{\delta}(p_3) \right]  
		\left [ - g_{\tilde{X}_1} \varepsilon_{\mu_1\tau\rho\phi}p_4^{\mu_1}\epsilon^\tau(p_4) \right]\nonumber\\
		&&\times\left [ \frac{-g^{\mu \alpha} +p_1^\mu p_1^\alpha / m_1^2}{p_1^2-m_1^2}  \right]
		\left [ \frac{-g^{\nu \rho} +p_2^\nu p_2^\rho / m_2^2}{p_2^2-m_2^2}  \right ] \left [ \frac{-g^{\beta \phi} +q^\beta q^\phi / m_q^2}{q^2-m_q^2}  \right ] \mathcal{F}^2\left ( q^2, m_q^2\right )\nonumber\\
  		\mathcal{M}_{B^+ \to D^{*+}\tilde{X}_2}^{(a)} &=& i^3\int \frac{d^4 q}{(2\pi)^4} \left[\mathcal{A}_{\nu } ^{B\to D_{s0}\bar{D}^*} (p_1,p_2)\right] \left[ i g_{D_{0}D^*V}  g_{\delta \beta} \epsilon^{\delta}(p_3) \right]  
		\left [ i g_{\tilde{X}_2} \epsilon_{\rho\phi}(p_4) \right]  \nonumber\\
		&&\times \left [ \frac{1}{p_1^2-m_1^2}  \right]
		\left [ \frac{-g^{\nu \rho} +p_2^\nu p_2^\rho / m_2^2}{p_2^2-m_2^2}  \right ] \left [ \frac{-g^{\beta \phi} +q^\beta q^\phi / m_q^2}{q^2-m_q^2}  \right ] \mathcal{F}^2\left ( q^2, m_q^2\right )\nonumber\\
		\mathcal{M}_{B^+ \to D^{*+}\tilde{X}_2}^{(b)} &=& i^3\int \frac{d^4 q}{(2\pi)^4} \left[\mathcal{A}_{\mu\nu } ^{B\to D_{s1}\bar{D}^*} (p_1,p_2)\right] \left[ i g_{D_{1}D^*V}  \varepsilon_{\delta\beta\alpha\beta_2}(p_1+p_3)^{\beta_2} \epsilon^{\delta}(p_3) \right]  
		\left [ i g_{\tilde{X}_2} \epsilon_{\rho\phi}(p_4) \right]\nonumber\\
		&&\times \left [ \frac{-g^{\mu \alpha} +p_1^\mu p_1^\alpha / m_1^2}{p_1^2-m_1^2}  \right]
		\left [ \frac{-g^{\nu \rho} +p_2^\nu p_2^\rho / m_2^2}{p_2^2-m_2^2}  \right ] \left [ \frac{-g^{\beta \phi} +q^\beta q^\phi / m_q^2}{q^2-m_q^2}  \right ] \mathcal{F}^2\left ( q^2, m_q^2\right )\nonumber
	\end{eqnarray}
\end{widetext}

\end{document}